\documentclass[journal=jacsat,manuscript=article, natbib]{achemso}
\usepackage{xcolor}
\usepackage{array}
\usepackage{upgreek}
\usepackage{amsthm}
\usepackage{mathtools}
\usepackage{physics}
\usepackage{amssymb}
\usepackage[version=3]{mhchem} 
\usepackage{soul}
\usepackage{hyperref}
\usepackage{amsmath}

\DeclareUnicodeCharacter{2212}{-}
\usepackage{comment}

\author{Ankita Karmakar}
\affiliation{Department of Physics, Indian Institute of Technology Kharagpur, Kharagpur 721302, India}
\author{Abhishek Mandal}
\affiliation{Department of Physics, Indian Institute of Technology Kharagpur, Kharagpur 721302, India}
\author{Maruthi M. Brundavanam}
\email{bmmanoj@phy.iitkgp.ac.in}
\affiliation{Department of Physics, Indian Institute of Technology Kharagpur, Kharagpur 721302, India}

\title{Engineering of Tunable Topological Texture Transformation in Optical Skyrmions and Bimerons using Enantiomeric Excess}

\begin{document}
\begin{spacing}{1.2}
\begin{abstract}
Optical skyrmions, which are the topologically protected quasiparticles and characterized by the nontrivial polarization textures, have emerged as a promising candidate due to their potential applications in optical communication, data storage, and particle manipulation. In this article, we propose and experimentally demonstrate an efficient and tunable approach for the dynamic transformation of generalized optical skyrmionic textures through the interaction of structured vector vortex beams with chiral media. By controlling the enantiomeric excess of an optically active material, we achieve on-demand conversion among Bloch, Neel or any intermediate skyrmionic states, extending also to optical bimerons. The topological conservation of the skyrmion number proves its robustness towards even higher-order textures. While maintaining a common path and stable setup, the proposed methodology provides an efficient and cost-effective approach towards the flexible manipulation of the topological textures, paving the way towards the understanding of topological transformation and engineering optical skyrmions for information processing or particle manipulation.

\end{abstract}
\section{Introduction}
Skyrmions, the topologically protected quasiparticles characterized by nontrivial spin or polarization textures, were first proposed by particle physicist Tony Skyrme in 1962 in the field of classical field theory \cite{skyrme1962unified}. 
The distinctive and topologically preserved characteristic has enabled new insights towards fundamental studies in condensed matter systems \cite{al2001skyrmions, muhlbauer2009skyrmion, bogdanov2020physical}, classical \cite{wang2025topological}, and quantum fields \cite{ornelas2024non} due to their remarkable robustness against external perturbations and defects. Recently, the photonic counterpart of the magnetic skyrmions has been observed first in structured electric fields of evanescent waves \cite{tsesses2018optical}, followed by spin vectors of confined free space waves \cite{du2019deep}, Stokes' vectors in paraxial vector beams \cite{PhysRevA.102.053513, lin2021microcavity}, pseudospins in photonic crystals \cite{guo2020meron, 9961875} and other new forms \cite{wang2024topologicals}. The advancement in the study of optical skyrmions has found promising application in deep subwavelength microscopy \cite{du2019deep}, precision particle manipulation \cite{lin2021photonic}, optical storage \cite{wang2026storage}, optical communication \cite{he2024optical}, etc. The topological nature of the optical skyrmions has proven its robustness against various perturbations, such as atmospheric turbulence \cite{wang2025robustness}, complex media \cite{wang2024topological}, for the aid of high-density data applications like optical communications and photonic computing. On the other hand, the dynamical transformation between topological textures has found significant importance in the optical transfer of information. Recently, the generation and topological transformation of Stokes' skyrmion between different textures have been realised leveraging the digital hologram method with a spatial light modulator \cite{shen2022generation}, polarization devices \cite{teng2023physical}, or induced by Gouy-phase \cite{chen2025gouy}. The dynamical manipulation of different skyrmionic textures has recently been studied using the Pancharatnam-Berry phase \cite {liu2025control}, utilizing a combination of wave-plates placed in specific positions of the Mach-Zehnder interferometer. Despite substantial progress in generating optical skyrmions, the dynamic and controllable transformation of their topological textures remains an important challenge. Existing approaches primarily focus on static generation or require modifications of the optical architecture, limiting their adaptability for reconfigurable photonic applications. In particular, achieving on-demand switching among Bloch \cite{gilbert2015realization}, Neel \cite{kezsmarki2015neel}, and intermediate skyrmionic textures within a stable experimental platform remains largely unexplored. A tunable mechanism capable of dynamically reconfiguring these topological states while preserving their topological protection is therefore highly desirable. Chiral media \cite{mason1982molecular} provide an effective and versatile route for tailoring structured light due to their intrinsic ability to differentially interact with polarization states \cite{karmakar2026enantiomeric}. The enantiomeric composition of optically active materials introduces a controllable chiral response \cite{mason1982molecular}, enabling precise manipulation of polarization textures without requiring complex modifications to the optical geometry. This opens the possibility of exploiting chirality as an active control parameter for engineering topological transformations in optical skyrmionic systems. \\
\indent In this article, a common-path and efficient methodology is proposed for the dynamical transformation of the Stokes' skyrmionic \cite{ma2025tailoring} textures leveraging enantiomeric excess ($EE$) \cite{karmakar2026enantiomeric} of an optically active (OA)  material. The control over the $EE$ provides the flexibility in tuning the helicity of the Stokes' vectors that defines the different topological textures. The proposed approach offers a fine level of tunability in a stable and cost-effective way, eliminating complex holography or propagation-assisted effects. The topological robustness is also proved using the preserved nature of the skyrmion number. The beauty and success of the tunable transformation of optical skyrmion have inspired us to extend the exploration towards the generalized topological texture, called optical bimeron \cite{shen2021topological}. Bimerons are quasiparticles that are homeomorphic to skyrmions and have potential applications in advanced information processing, transport, and optical storage \cite{PhysRevB.99.060407,jani2021antiferromagnetic, zhang2021frustrated} etc. The topological manipulation of optical bimerons is also shown to be achieved using a birefringent waveplate, along with the control over $EE$, which can be useful in chiral separation and sorting of chiral particles\cite{xue2026chiral}.\\
\indent In the present work, the Stokes' skyrmions are generated using the vector vortex beams (VVBs) in a common-path technique. The VVBs form the skyrmionic structure, which is transformed to any arbitrary desired skyrmionic texture by suitably choosing the $EE$ of the OA solution. The tuable topological transformation scheme is extended to a more generalized form, called the bimeron. The preservation of the skyrmion number also proves the robustness of the proposed methodology, which is also valid for higher-order topological textures. For a better visual interpretation, the generalised skyrmions are represented on a generalised skyrmion torus, where different parameters define the different types of textures. The proposed method provides a stable, flexible, and tunable way to manipulate the arbitrary topological textures with profound applications in optical communication and quantum information.

\section{Theoretical Details}
\subsection{Topological description of skyrmions}
The topological properties of the optical skyrmions can be characterized in terms of skyrmionic number \cite{nagaosa2013topological} given as,
\begin{equation}
    S=\dfrac{1}{4\pi}\iint\limits_{\sigma} \textbf{n}\cdot\left(\frac{\partial \textbf{n}}{\partial x}\times\frac{\partial \textbf{n}}{\partial y} \right ) \,dx\,dy
\end{equation}
where $\textbf{n}(x,y)$= \textbf{n}($r \cos \phi$, $r \sin \phi$) represents the vector field to construct the quasiparticle confined in a region $\sigma$. The skyrmion number is an integer that denotes the number of times the vector \textbf{n} wraps the unit sphere. For the wrapping of the unit sphere, the vector \textbf{n} can be written in terms of the azimuth and ellipticity angle, $\alpha(\phi)$ and $\beta(r)$, respectively, as,
\textbf{n}=$\left(\cos \alpha(\phi)\sin\beta(r), \sin\alpha(\phi)\sin\beta(r), \cos\beta(r)\right)$ and the skyrmion number can be separated into two additional topological number as, 
\begin{equation}
\begin{split}
    S&=\dfrac{1}{4\pi}\int\limits_{0}^{r_{\sigma}}\,dr\int\limits_{0}^{2\pi}\,d\theta\dfrac{\,d\beta(r)}{\,dr}\dfrac{\,d\alpha(\theta)}{\,d\theta}\sin \beta(r)\\
    &=\dfrac{1}{4\pi}\left[\cos \beta(r)\right]_{r=0}^{r=r_\sigma}\left[\alpha(\theta)\right]_{\theta=0}^{\theta=2\pi}\\
    &=p.m.
    \end{split}
\end{equation}
Here, the polarity $p=\dfrac{1}{2}\left[\cos\beta(r)\right]_{r=0}^{r=r_\sigma}$ and the vorticity $m=\dfrac{1}{2\pi}\left[\alpha(\theta)\right]_{\theta=0}^{\theta=2\pi}$ are the two fundamental topological parameters that defines the structural properties of the skyrmion. The polarity defines the out-of-plane vector direction down (up) at the center $r=0$ and up (down) at the boundary $r=r_\sigma$ for $p=1$$(p=-1)$. The parameter $m$ defines the transverse vector orientation. For characterizing the distinct helical structure of the skyrmion, an initial phase is added as: $\alpha(\theta)=m\theta+\gamma$. For a specific latitude angle $\beta$, the initial phase $\gamma$ defines the inclination angle of the initial vector in a circular way. For a given value of $m$, the skyrmions can have different topological textures depending on the initial phase $\gamma$. For example, when $m=1$, the skyrmionic texture is of Neel-type for $\gamma=k\pi$ (k=integer), resulting in a hedgehog-like structure. Similarly, a vortex profile corresponding to $\gamma=(2k+1)\pi/2$ is called the Bloch-type skyrmion. The saddle-like texture corresponding to $m=-1$ is always called an anti-skyrmion. The diverse topological characteristics of the skyrmions are governed by the $\gamma$.\\
\indent In free space, the optical skyrmions can be constructed from the Stokes' vectors, called the Stokes' skyrmion \cite{ma2025tailoring}. They are generated from the structured vector field with non-uniform polarization distribution, known as the vector vortex beams (VVBs) \cite{karmakar2026enantiomeric, patra2025tuning}. They are realized by the superposition of the Gaussian and a vortex beam with orthogonal polarization and are represented as,
\begin{equation}
    \psi = LG_0^0\hat{\textbf{e}}_0+e^{i\gamma}LG_p^l \hat{\textbf{e}}_1,
\end{equation}
where, $\hat{\textbf{e}}_0$ and $\hat{\textbf{e}}_1$ are the two orthogonal polarizations (in this work, the orthogonal circular polarization state is chosen) and $LG_p^l$ represents the Laguerre-Gaussian mode \cite{lian2022oam} with radial index $p$ ($p=0$ is taken in this work) and azimuthal index $l$, also called the topological charge (TC). The relative phase difference between the two superposing beams, $\gamma$ defines the polarization distribution as well as the helicity configuration of the resulting skyrmion. \\
\indent The spatial distribution of polarization of the VVBs and thus the optical skyrmionic texture can be tuned by changing the relative phase difference between two orthogonal circular polarizations. The initial phase $\gamma$ can be tailored using the phenomenon of optical rotation \cite{haynes2016crc}, which is caused by different transmission speed experienced by left-handed and right-handed circularly polarized (LCP and RCP, respectively) light with the interaction of a chiral medium \cite{karmakar2024controlled, karmakar2026single, karmakar2026enantiomeric}. The RCP and LCP suffer from different transmission speeds with the interaction with chiral materials, whereas their opposite nature of interaction with dextro and levo component results in optical rotation (OR) in right-handed or left-handed, respectively. Chiral enantiomers are optically active substances that are non-superimposable to their mirror images. When two such enantiomers are mixed, that will cause an effective optical rotation of the plane-polarized light depending on the ratio of isomers. A homogeneous mixture of equal amounts of isomers, called the racemic mixture (50:50), results in opposite but equal transmission speeds of RCP and LCP, causing no effective optical rotation. When the VVBs interact with optically active enantiomers, the dextro or levo compounds introduce an additional phase to RCP and LCP, denoted as $\phi_1$ and $\phi_2$. Due to the interaction, the resultant VVBs will have the form,
\begin{equation}
     \psi= LG^{0}_{0}e^{-i\phi_1}\hat{e}_R+ LG^{1}_{0}e^{i(\gamma+\phi_2)}\hat{e}_L
     \label{eq3}
\end{equation}
which results in an additional intermodal phase difference of $\Delta\phi=\phi_2-\phi_1$, causing a net rotation of the linear polarization by an angle $\Delta\phi/2$. \\
\indent The OR \cite{karmakar2024controlled, karmakar2026single, karmakar2026enantiomeric} $\Delta\phi/2$ = $\Theta$ experienced by a linearly polarized light due to the interaction with a chiral material is given by,
\begin{equation}
    \Theta=l\times c\times [s]_\lambda^T,
    \label{eq4}
\end{equation}
where $\textit{l}$ is the path length of the light through the material, $\textit {c} $ is the concentration, and $[s]_{\lambda} ^T$ is the specific rotation of the chiral material at a particular temperature and wavelength of light. Similarly, the optical rotation $\Theta_{mix}$ of a mixture of two enantiomers (dextro and levo) is given by, 
\begin{equation}
\begin{split}
    \Theta_{mix}= l\times (c_d\times s+c_l\times(-s))
                 =l\times (c_d-c_l)\times s.
    \end{split}
    \label{eq5}
\end{equation}
Here, $s$ and $-s$ define the same, but opposite-handed specific rotation (SR) of the dextro (d) and levo (l) enantiomer, and $c_d$ and $c_l$ are the concentrations of the d and l enantiomers, respectively, following $c_d+c_l=c$. The excess amount of one enantiomer over the other is called the enantiomeric excess (EE) and defined as $\dfrac{c_d-c_l}{(c_d+c_l)}\times 100 \% $. \\
\indent To incorporate the effect of the OR in controlling the skyrmionic structures, the azimuthal phase variation is modified to \begin{equation}
    \alpha(\theta)=m\theta+\gamma+\gamma_{OA},
\end{equation}
where, the $\gamma_{OA}$ is tuned by controlling the EE of the enantiomeric mixture, resulting in a dynamic topological transformations between different skyrmionic textures.\\
\indent Optical bimeron, being a generalized topologically transformed state of optical skyrmion, consists of two half skyrmions (meron), with opposite polarity (one half-skyrmion and one half anti-skyrmionic texture, respectively). The combination of two merons form the topological bimeron structure, which is constructed by the superposition of horizontal and vertically polarized beams with a relative phase difference between them. The optical skyrmions can be topologically transformed into an optical bimeron using a birefringent waveplate, and the different topological texture of the bimeron can be dynamically tuned using different EE of the enantiomeric mixture.

\section{Experimental setup}
\begin{figure*}[h!]
 \centering
    \includegraphics[scale=0.42]{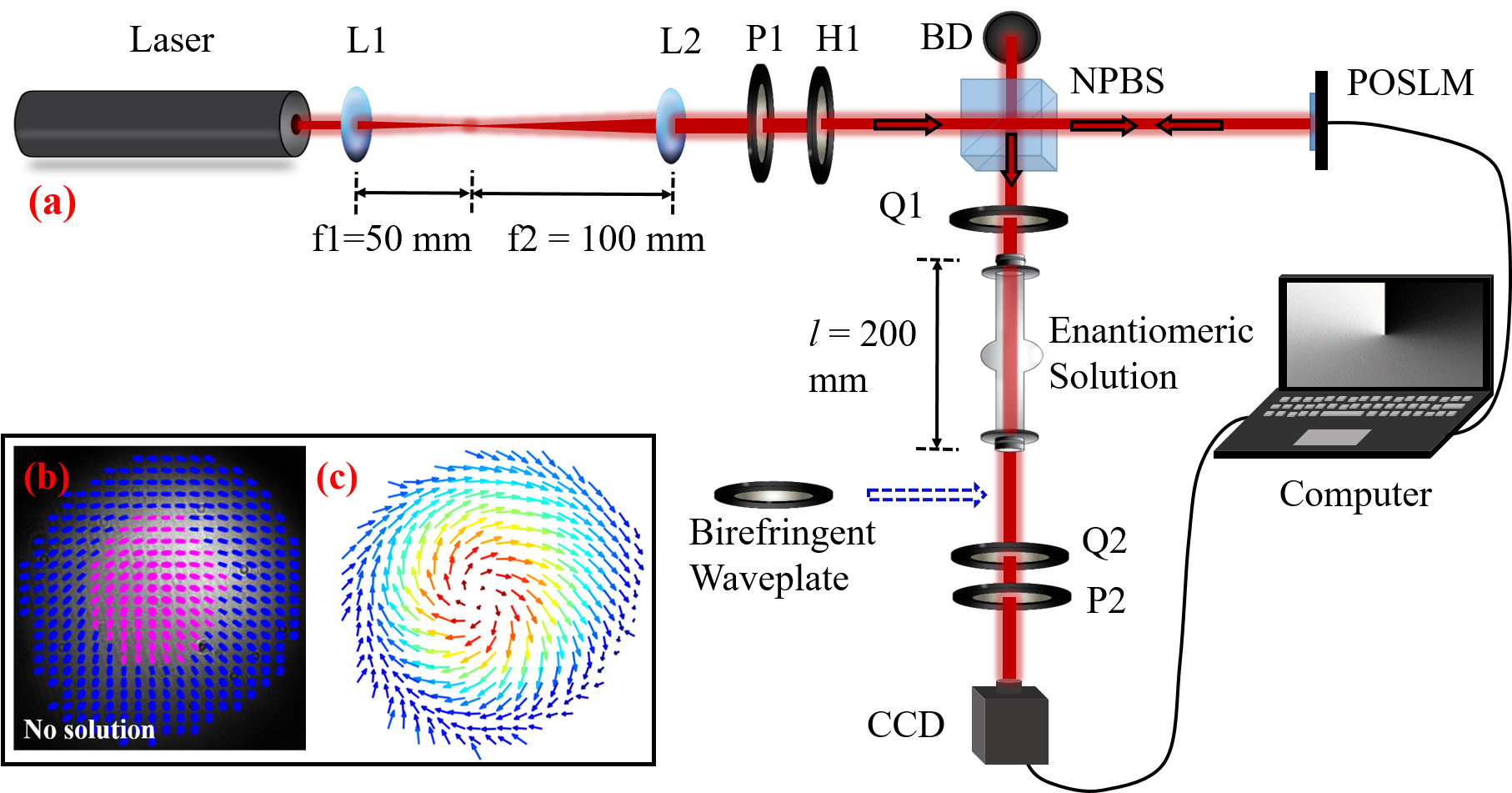}
   \caption{Schematic diagram of the experimental setup for the tunable transformation of Stokes' skyrmions using enantiomeric excess; L1, L2: Lenses; P1, P2: Polarizers; H1: Half wave plate; NPBS: Non-polarizing beam splitter; BD: Beam dump; Q1, Q2: Quarter wave plates; POSLM: Phase only spatial light modulator; CCD: Charge coupled device; (b) Spatial polarization distribution of the vector vortex beams before the optically active solution and (c) its corresponding Bloch-type skyrmionic texture.}
    \label{Fig1}
   \end{figure*}
The optical skyrmions are constructed from the VVBs as shown in Fig. 1(a). The Gaussian laser beam with wavelength 633 nm is collimated and expanded using a two-lens system L1 and L2 with focal lengths f1 and f2 of 50 mm and 100 mm, respectively. The collimated beam is passed through a polarizer and a half-wave plate (HWP) H1 with its fast axis placed at an angle $22.5^\circ$ to convert the beam into a diagonally polarized beam. The beam is directed to the POSLM (Holoeye Pluto 2.1, 1920$\times 1080$, 8 $\mu m$ pixel pitch) through a non-polarizing beam splitter (NPBS). The POSLM modifies the x-polarization component with the vorex profile of unit TC whereas, whereas the y-polarized component remains unaltered. The superposed beam is then directed in an orthogonal direction through the NPBS and passed though a quarter-wave plate (QWP) Q1 with its fast axis at $-45^\circ$, which converts the two orthogonal linear polarized beams into two orthogonal circular polarizations having different TCs, resulting in VVB. The generated VVB is sent through the OA enantiomeric mixture with variable EE. A relative phase difference between two superposing orthogonal circular polarizations ($\delta_{OA}$) is introduced depending on the value of EE, resulting in a modified polarization structure and a different VVB. For the characterization of the generated VVBs for different EEs, the intensities corresponding to six different polarization projections (horizontal-H, vertical-V, diagonal-D, anti diagonal-A, right circular-R, and left circular-L) are recorded with a charge-coupled device (CCD) (1280 $\times$ 1024, 4.65 $\mu m$ pixel pitch). The Stokes' parameters \cite{kihara2011measurement} are estimated from the recorded intensity patterns as,
\begin{equation}\begin{split}
  &S_0= I_H+I_V\\ 
  &S_1=I_H-I_V\\
  &S_2=I_D-I_A\\
  &S_3=I_R-I_L.
\end{split}
\end{equation}
The polarization ellipse parameters are estimated using the Stokes' parameters, and the polarization distribution of the VVBs is reconstructed over the beam profile. The VVBs are represented in terms of the optical skyrmion, where different polarization structures result in different skyrmionic structures, which are controlled using the variation of the EE.\\
\indent The optical skyrmions are topologically transformed into a generalized skyrmionic state, an optical bimeron using a birefringent waveplate. A QWP with its fast axis at $-45^\circ$ with respect to the x-axis is used to convert the orthogonal circular basis to the horizontal and vertical basis. The bimerons are also dynamically transformed from one topological texture to another by controlling the EE of the enantiomeric mixture, where the relative phase difference between the superposing polarizations is controlled by the QWP together with the varying EE of the medium.

\section{Results and Discussions}
\subsection{1. Tunable Topological Texture}
\begin{figure*}[h!]
 \centering
    \includegraphics[scale=0.39]{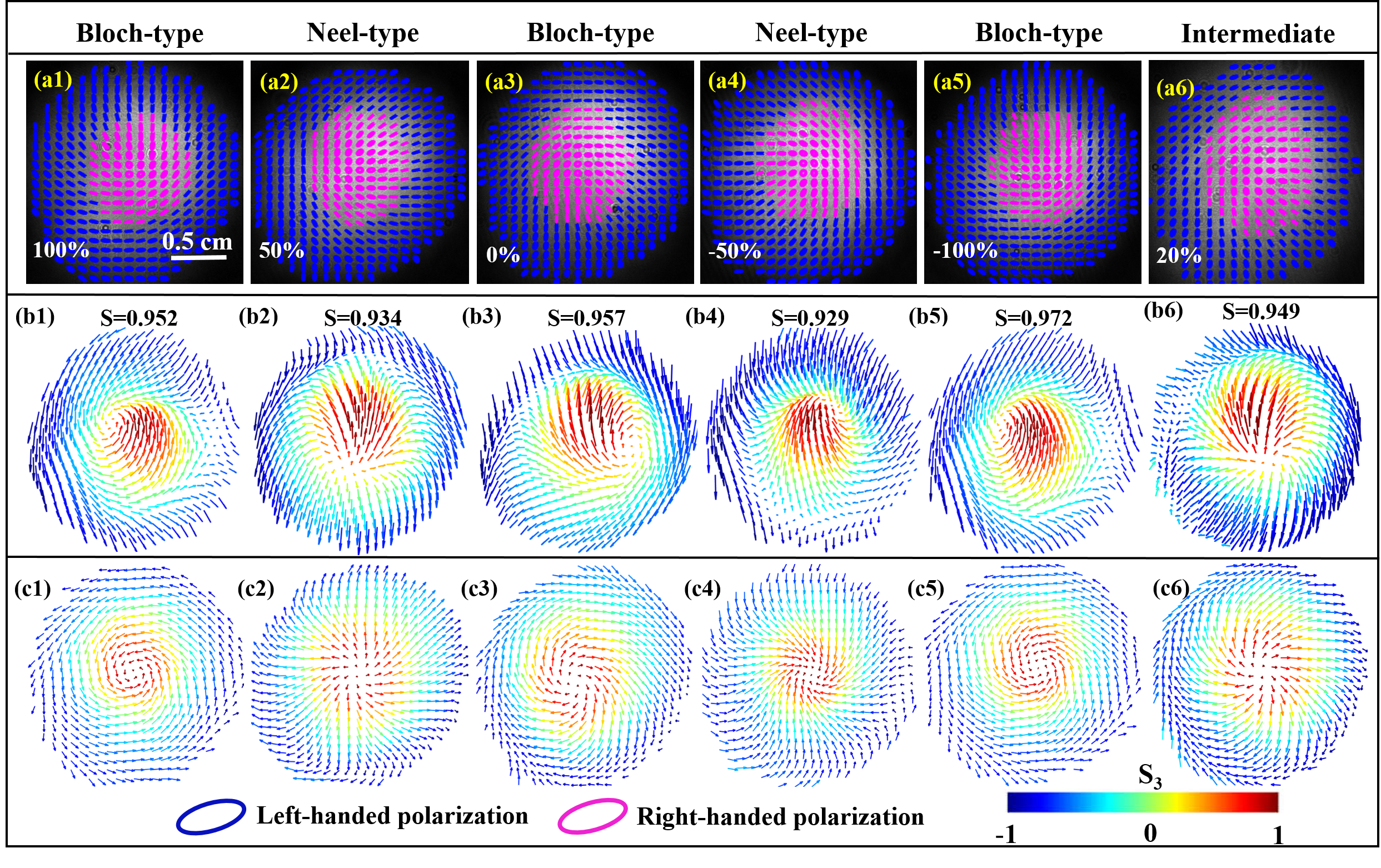}
   \caption{(a1-a6) Spatial distribution of polarization for the skyrmionic vector vortex beams generated with different enantiomeric excess (mentioned in the left-bottom corner); (b1-b6) The corresponding Stokes' skyrmionic texture with the vector in 3D representation, and (c1-c6) transverse component vector. The topological texture corresponding to each $EE$ is mentioned at the top of each column.}
    \label{Fig1}
   \end{figure*}
\noindent The optical skyrmions and bimerons constructed from the VVBs are generated as shown and described using the setup as shown in Fig. 1(a). In the absence of the OA material, the generated VVB (shown in Fig. 1(b)) is a superposed state of the Gaussian and OV beam with orthogonal circular polarization, thus forming a skyrmion with a vortex core at the center, which defines its Bloch-type nature as shown in Fig. 1(c). To generate a Neel-type texture, an extra relative phase difference of $\gamma_{OA}=\pm\pi/2$ is to be introduced, whereas, for a Bloch-type, the requirement of phase difference is $\gamma_{OA}=\pm\pi$. Due to the interaction of the VVB with an enantiomeric mixture of varying EE, the resultant polarization structure of the VVB changes, resulting in a different topological skyrmionic texture. Here, (+)-limonene is chosen as the pure enantiomer, and (-)-limonene is the impurity to it. The excessive amount of (+)-limonene is termed the EE in this article. The SR of the pure (+) and (-)-limonene are measured to be $84.25^\circ$ and $-84.25^\circ$, respectively, at room temperature and 633 nm wavelength. The concentration of the solution is chosen 
\begin{figure*}[h!]
 \centering
    \includegraphics[scale=0.51]{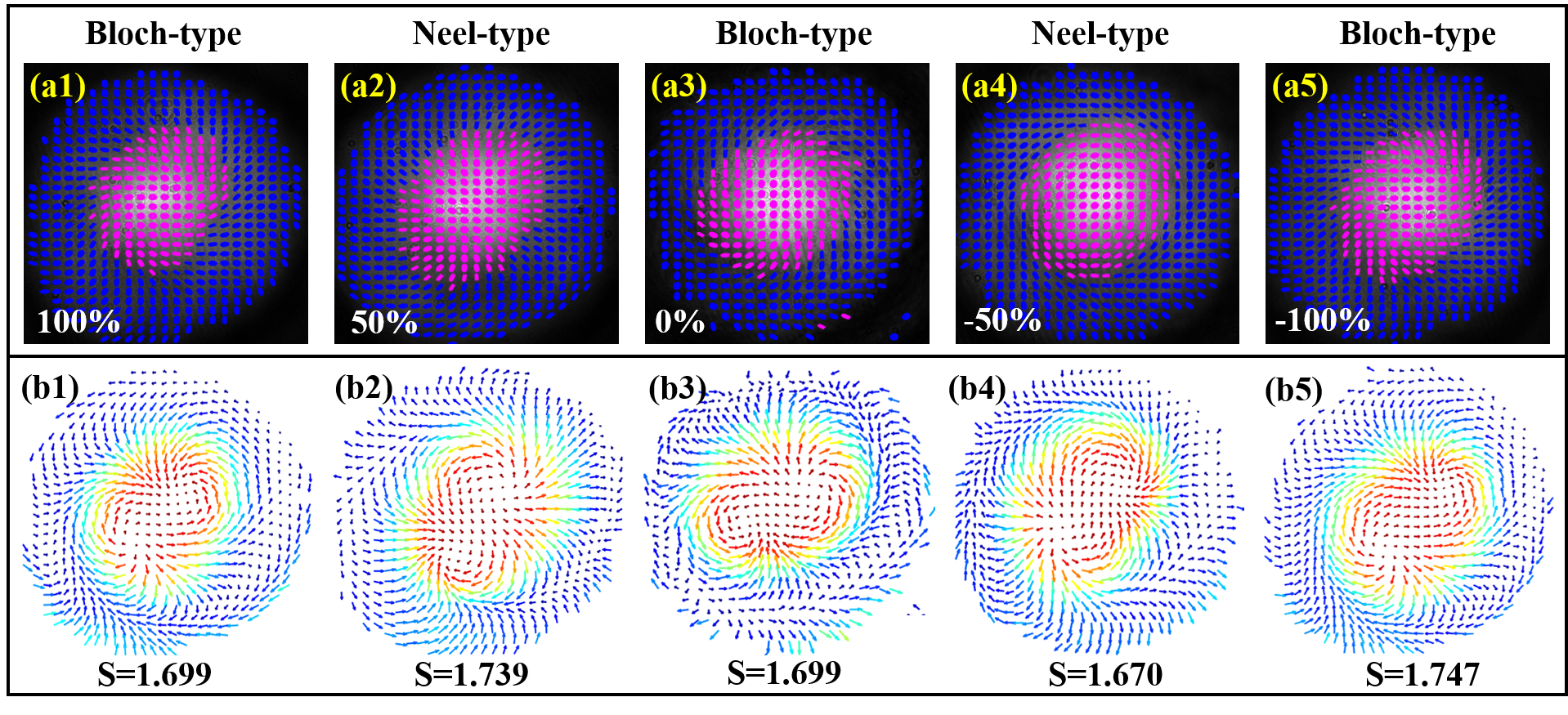}
   \caption{(a1-a5) Spatial distribution of polarization for the second-order skyrmionic beam generated with different enantiomeric excess (mentioned in the left-bottom corner); (b1-b5) The corresponding Stokes' skyrmionic texture with the transverse vector in 2D representation. The topological texture corresponding to each $EE$ is mentioned at the top of each column.}
    \label{Fig1}
   \end{figure*}
accordingly as $c=53.4\%$ that provides an OR of $\pm90^\circ$ for an $EE=\pm100\%$ at the same temperature and wavelength. The initial VVB representing a Bloch-type skyrmion interacts with the solution with $EE $ in the range from $-100\%$ to $100\%$, and the corresponding polarization map for the generated VVBs is shown in the first row of Fig. 2. The second row and the third row of Fig. 2 represent the corresponding Stokes' vector in the form of 3D representation and their transverse vector component in 2D. An $EE=100\%$ of the solution introduces $\gamma_{OA}=+\pi$, resulting in a Bloch-type skyrmion as shown in Fig. 2(a1-c1). When the $EE$ is reduced to $+50\%$, the $\gamma_{OA}=\pi/2$ generates a Neel-type skyrmion with a hedgehog texture as can be seen in Fig. 2(a2-c2). If the mixture consists of equal amounts of (+) and (-)-limonene, the resultant OR as well as the $\gamma_{OA}=0$, resulting in an unaltered state of polarization, resulting in the Bloch-type skyrmion that is alike the initial state, as clearly visible from Fig. 2(a3-c3). In a similar manner, an $EE=-50\%$ and $EE=-100\%$ generate the Neel-type and Bloch-type texture of skyrmions as visible from Fig. 2(a4-c4) and 2(a5-c5), respectively. The skyrmion number estimated using Eq. 1, corresponding to each texture (shown in Fig. 2), also confirms the topological stability during the transformation of the skyrmionic textures. The control over the $EE$ of the enantiomeric mixture can dynamically transform from one topological texture to any desired intermediate state (as shown in Fig. 2(a6-c6)) without the aid of any other optical conversion. In a similar manner, the second-order (as shown in Fig. 3) or higher-order skyrmions can also be transformed dynamically using different $EE$. The tunable transformation of the second-order skyrmions is shown in Fig. 3 in terms of the reconstructed polarization structure and their transverse vector representation. The estimated skyrmion number also proves the topological robustness during the transformation of the texture.\\
\begin{figure*}[h!]
 \centering
    \includegraphics[scale=0.39]{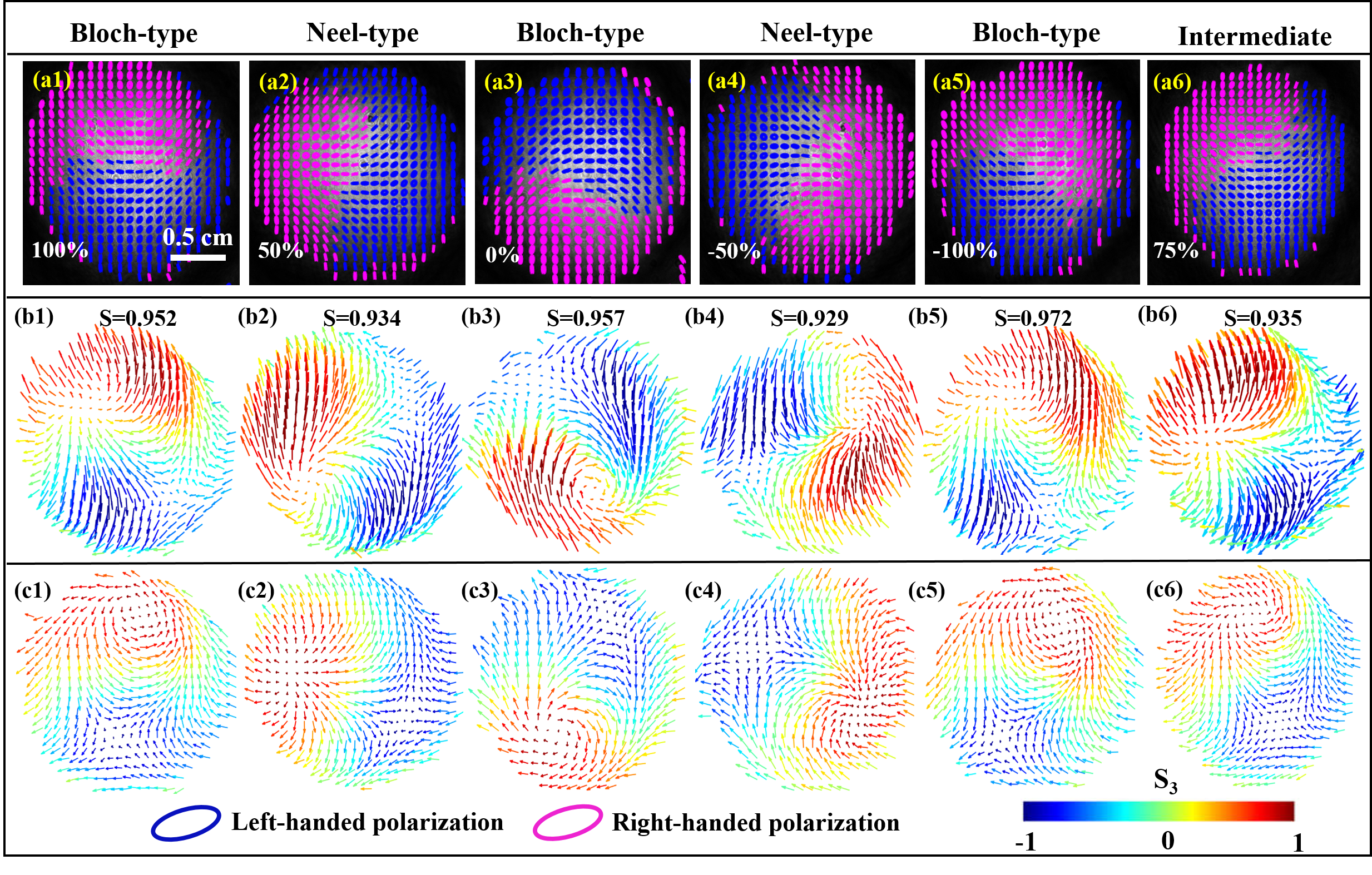 }
   \caption{(a1-a6) Spatial distribution of polarization for the bimeronic vector vortex beams generated with different enantiomeric excess (mentioned in the left-bottom corner); (b1-b6) The corresponding Stokes' bimeronic texture with the vector in 3D representation, and (c1-c6) transverse vector component. The topological texture corresponding to each $EE$ is mentioned at the top of each column.}
    \label{Fig1}
   \end{figure*}
\indent The skyrmions are converted into optical bimerons using a QWP, and the $\delta_{OA}$ between the two interfering orthogonal linear polarizations (horizontal and vertical) is controlled using the $EE$ of the enantiomeric mixture. In the absence of the OA medium, the generated texture consists of one vortex and one saddle texture \cite{shen2021topological}, resulting in a Bloch-type bimeron. When the VVBs are passed through the OA medium and the $EE=100\%$, due to the introduction of  $\gamma_{OA}=\pi$, the resulting texture again forms a Bloch-type bimeron as shown in Fig. 4(a1-c1). When the $EE$ is reduced to $50\%$, the resultant bimeron is converted to a Neel-type which has one hedgehog and one saddle texture \cite{shen2021topological} as shown in Fig. 4(a2-c2). A racemic mixture will not have any effect on the texture, thus remains Bloch-type bimeron (shown in Fig. 4(a3-c3)), but again for a $EE=-50\% $ and $EE=-100\%$, the resulting textures are Neel-type and Bloch-type (shown in Fig. 4(a4-c4) and 4(a5-c5), respectively). The $EE$ can transform the bimeronic texture to any other desired intermediate texture as shown in Fig. 4(a6-c6). The estimated skyrmion number in the case of bimeron also matches with the value of the TC. The dynamic transformation of optical bimeron is also valid and applicable for higher-order texture, as shown for second-order bimeron in Fig. 5.\\
\begin{figure*}[h!]
 \centering
    \includegraphics[scale=0.51]{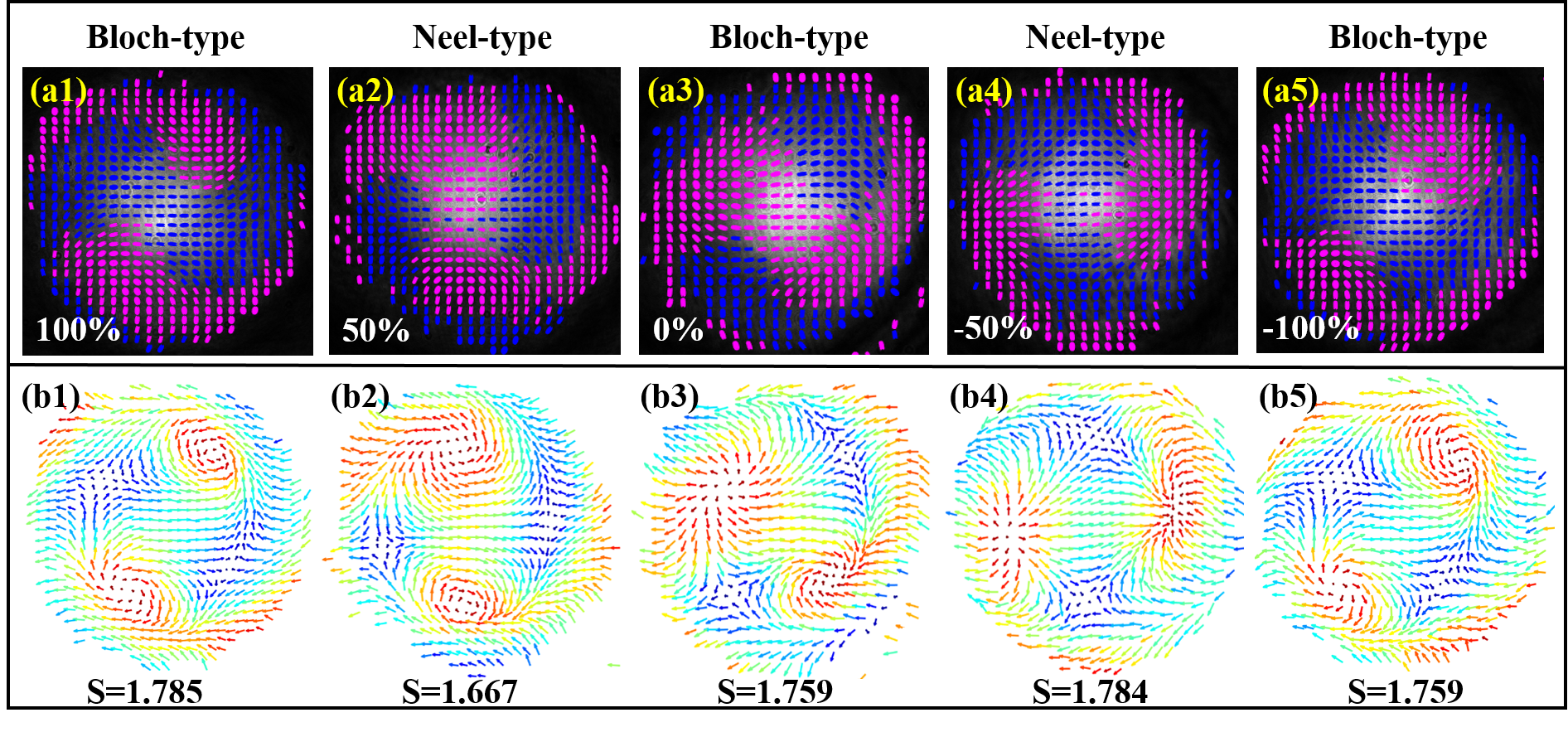}
   \caption{(a1-a5) Spatial distribution of polarization for the second-order bimeronic beam generated with different enantiomeric excess (mentioned in the left-bottom corner); (b1-b5) The corresponding Stokes' bimeronic texture with the transverse vector in 2D representation. The topological texture corresponding to each $EE$ is mentioned at the top of each column.}
    \label{Fig1}
   \end{figure*}

\subsection{2. Generalized Skyrmion Torus}

\begin{figure*}[h!]
 \centering
    \includegraphics[scale=0.46]{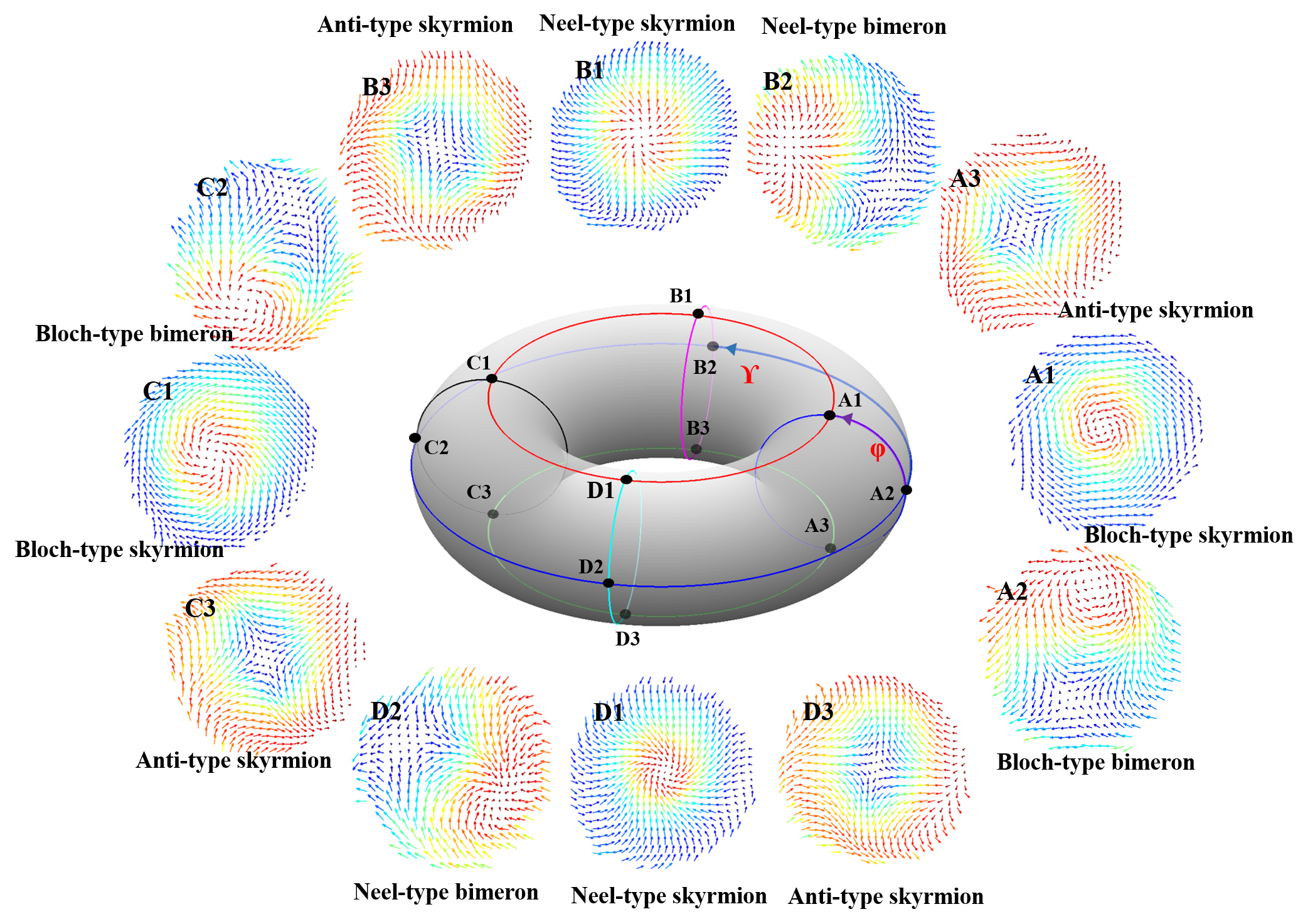 }
   \caption{Generalized skyrmion torus shown for the tunable transformation of generalized skyrmionic textures using the enantiomeric excess. The polar angle $\phi$=$0,\pi$ and $3\pi/2$ represents the bimeron, skyrmion, and anti-skyrmionic state, whereas the toroidal angle $\gamma$ (Ai-Di)(i=1,2,3) corresponds to Neel, Bloch, and any intermediate state that can be controlled by $ EE$}.
    \label{Fig1}
   \end{figure*}
For the deeper insight of the dynamical transformation of the optical skyrmions and bimerons, the concept of the generalized skyrmion torus is presented, where each point on the torus represents a generalized optical skyrmion. The two parameters $\gamma$ and $\phi$, respectively the toroidal and polar angles of the torus, represent the optical skyrmion and optical bimeron family for particular cases. In Fig. 6, the first-order generalized skyrmion is presented. For $S>0$, $\phi=0$ represents the bimeron plane, which is the big outer circle (shown in blue color in Fig. 6, represented by points A2-D2) of the torus, whereas $\phi=+\pi/2$ represents the skyrmion family (shown in red color, A1-D1) and $\phi=3\pi/2$ is the anti-skyrmion family (shown in green color, A3-D3). The toroidal angle $\gamma (\gamma_{OA})$ defines the different types of skyrmionic (bimeronic) textures. $\gamma_{OA}=0$ or $\pi$ represents Bloch-type texture wheras, $\gamma_{OA}=\pm\pi/2$ represents Neel-type texture. The anti-skyrmion texture is also shown to be tunably transformed by controlling the $EE$ in a similar way to that of skyrmion. Having a precise and regulated control over the $EE$ of the enantiomeric mixture for a suitable concentration, the parameter $\gamma_{OA}$ can be adjusted to dynamically transform any generalized skyrmionic state to another.

\section{Conclusions}
In conclusion, we propose a stable, efficient, and common-path technique for the dynamic transformation of diverse topological textures of generalized optical skyrmions. By exploiting the interaction of VVBs with an enantiomeric mixture possessing controlled $EE$, the optical skyrmionic textures can be dynamically manipulated and transformed. The proposed approach enables not only the reversible conversion between Bloch-type and Neel-type skyrmions but also the tunable transformation of any skyrmionic state into another desired intermediate topological configuration. Furthermore, through the incorporation of a birefringent wave plate, optical bimeronic textures can also be flexibly converted by tailoring the $EE$ of the enantiomeric mixture. Importantly, the scheme remains effective for higher-order skyrmionic and bimeronic structures. The preservation of the skyrmion number during these dynamical transformations highlights the topological robustness of the proposed approach. By reducing the need for multiple optical wave plates and utilizing an OA enantiomeric mixture with controlled EE, the present framework offers a versatile platform for the dynamic engineering of generalized optical quasiparticles like skyrmions and bimerons with promising implications for advanced optical and quantum communication technologies.

\section*{Acknowledgement}
A. Karmakar gratefully
acknowledges IIT Kharagpur for the fellowship. Maruthi M. Brundavanam thanks SERB (Grant no.: CRG/2019/006799) for financial assistance.



\bibliography{main}
\end{spacing}
\end{document}